\documentclass[pra]{revtex4-2}
\usepackage{braket}
\usepackage{graphicx}

\begin{document}

\title{Uncertainty limits of the information exchange\\ between a quantum system and an external meter}

\author{Tomonori Matsushita and Holger F. Hofmann}
\email{hofmann@hiroshima-u.ac.jp}
\affiliation{Graduate School of Advanced Science and Engineering, Hiroshima University, Kagamiyama 1-3-1, Higashi Hiroshima 739-8530, Japan}


\begin{abstract}
It is not possible to obtain information about the observable properties of a quantum system without a physical interaction between the system and an external meter. This physical interaction is described by a unitary transformation of the joint quantum state of the system and the meter, which means that the information transfer from the system to the meter depends on the initial quantum coherence of the meter. In the present paper, we analyze the measurement interaction in terms of the changes of the meter state caused by the interaction with the system. The sensitivity of the meter can then be defined by evaluating the distinguishability of meter states for different values of the target observable. It is shown that the sensitivity of the meter requires quantum coherences in the generator observable that determines the magnitude of the back action of the meter on the system. The trade-off between measurement resolution and back action is decided by the relation between sensitivity and quantum coherent uncertainty in the external meter. No matter how macroscopic the device used as a meter is, it must necessarily be quantum coherent in the degrees of freedom that interact with the system being measured.
\end{abstract}

\maketitle

\section{Introduction}

Quantum measurements are the essential interface between a quantum system and its environment. It is the goal of a quantum measurement to efficiently access the information contained within the quantum system. The achievement of this goal is complicated by the uncertainty principle of quantum mechanics which requires a minimal disturbance of the system whenever the value of a physical property is determined \cite{Hei27}. The precise trade-off between the information obtained in a quantum measurement and the corresponding disturbance can be formulated in a surprisingly large number of ways, resulting in a variety of different approaches to this fundamental problem of quantum mechanics \cite{Oza03,Bus07,Bus13,Bus14,Bcm14,Rod19}. A major difficulty addressed by these approaches is the evaluation of information encoded in the statistics of incompatible measurements \cite{Hal04,Tol07,Dre10,Hal16,Pia16,The18,Nii18,Hof20,Mar20}. In a completely quantum mechanical theory, the relations between the outcomes of incompatible measurements are expressed by quantum coherences that can describe contradictory statements about the predicted outcomes of these measurements \cite{KS,Fra18,Bru18,Pro19}. The disturbance of the system by the measurement interaction represents the necessary loss of quantum coherence needed to avoid observable contradictions between the measurement results of a sequence of measurements on the same system \cite{Rod18}. Measurement uncertainties are therefore essential to the resolution of quantum paradoxes \cite{Opp10,Cab10,Ban13,Hof19}. However, measurement uncertainties are difficult to identify because measurement interactions entangle the system and the meter, resulting in a steering scenario where the quantum mechanics of the meter determines the relation between measurement information and decoherence \cite{Pat19,Wis07}. It is important to realize that this emergence of entanglement in the measurement interaction is an unavoidable characteristic of all quantum measurements. Any realistic description of a quantum measurement must necessarily include the quantum coherence of the meter, raising questions about the conventional reduction of measurements to the classical information contained in their final outcome \cite{vKam88,All13}. In the present paper, we address this problem by investigating the essential role of quantum coherences of the external meter in the exchange of information between the system and the meter.

The problem we focus on is the separation between the system and the meter in the measurement interaction. Even though the effects of the interaction entangle the system and the meter, the physics of the interaction can be understood in terms of conditional unitary transformations representing either the action of the system on the meter or the back action of the meter on the system \cite{Hof05}. Information about the value of the target observable $\hat{A}$ is encoded in the meter as an amount of change corresponding to the unknown parameter $\varphi_B$ of a unitary operation in the meter. The sensitivity of the meter depends on the possibility of determining this parameter in a single read out measurement. Using a method similar to the one developed in \cite{Pat19} we find that the resolution of two possible values of $\hat{A}$ is a well-defined function of the difference between these two values, resulting in a quantitative measure of resolution. The mathematical form of this measure corresponds to the Fisher information of the meter, conveniently establishing the uncertainty of the generator observable $\hat{B}$ of the conditional unitary in the meter as the upper limit of the sensitivity. By applying this uncertainty to the amount of conditional change in the system described by the back action, it is possible to show that the quantum uncertainty of the meter is responsible for supplying the precise amount of decoherence required whenever two states can be distinguished by their effects on the meter. By separately analyzing the role of the meter in the measurement interaction we can therefore show that the decoherence in the system invariably originates from a quantum coherent uncertainty in the meter. The universal validity of quantum mechanics thus ensures that no physical system can be sensitive to the value of a target observable $\hat{A}$ unless it is itself quantum mechanically uncertain at the microscopic level, and this quantum uncertainty will be an essential element in the description of the dynamics of the system-meter interaction. It should be emphasized that the approach presented in the following shows that the quantum mechanics of the meter plays a role that is independent of the system state. It is therefore impossible to construct a ``classical'' measurement where the result can be recorded without introducing a corresponding amount of quantum uncertainty from outside the system. This fundamental property of the environment could explain how the measurement context is imposed on a quantum system during the measurement interaction. All macroscopic systems have intrinsic quantum fluctuations at the microscopic level, and these quantum fluctuations ensure that any information transfer between a system and its environment conforms to the context dependent laws of quantum mechanics, regardless of whether the information transfer is described at the microscopic level or at the macroscopic level.

The rest of the paper is organized as follows. In Sec. \ref{sec:separate} we separate the meter response and the back-action by focusing on the force to each system. In Sec. \ref{sec:sensitivity} we analyze the meter sensitivity and shows the meter response requires the uncertainty of the generator $\hat{B}$. In Sec. \ref{sec:system} we show that the cause of decoherence in the system is the uncertainty of the target observable $\hat{B}$ and the relation between the meter sensitivity and decoherence in the system is discussed. In Sec. \ref{sec:context} the role of decoherence in the resolution of quantum paradoxes is considered and the contextuality of realities observed by external meters is discussed. Sec. \ref{sec:conclude} summarizes the results and concludes the paper.

\section{Separation of measurement signal and back action}
\label{sec:separate}

Any measurement of a target observable $\hat{A}$ of a physical system requires an interaction between the system $S$ and an external meter $M$, so that the value of the physical quantity $\hat{A}$ can be obtained from the response of the meter to the force exerted by the system. In the absence of additional information about the input state of the system, the ideal interaction Hamiltonian that generates a meter response that depends exclusively on the magnitude of $\hat{A}$ is proportional to the dyadic product of the two operators, $\hat{A}\otimes\hat{B}$. Any additional terms in the interaction Hamiltonian would result in meter responses that depend on physical properties of the system other than the target observable $\hat{A}$. It might be worth noting that such errors can be compensated in the meter readout when prior information about the input state is available, resulting in a more sophisticated state-dependent evaluation of measurement errors for completely arbitrary system-meter interactions \cite{Oza03,Hal04,Nii18,Hof21}. However, in the absence of such additional information deviations from the ideal bi-linear form of the interaction Hamiltonian would only introduce unnecessary measurement errors associated with the meter response to system properties other than $\hat{A}$. Since we are specifically interested in the transfer of information about the target observable $\hat{A}$ from the system to the meter we will focus on the meter response achieved by interaction Hamiltonians proportional to $\hat{A}\otimes\hat{B}$, where the magnitude of the meter response is exclusively determined by $\hat{A}$. It should be noted that the present discussion is still completely general with regards to the Hilbert spaces of the system and the meter and requires no additional assumptions about the possible target observables $\hat{A}$ or the possible meter generators $\hat{B}$.

Since the measurement must be completed within a finite amount of time, the Hamiltonian of the interaction will be time dependent and the total measurement interaction is obtained by the time integral of the Hamiltonian. The result is a unitary transformation describing the total effects of the interaction as
\begin{equation} 
\label{eq:interaction}
\hat{U}_{SM} = \exp \left( - \frac{i}{\hbar} g \hat{A} \otimes \hat{B} \right),
\end{equation}
where $g$ describes the strength of the time-integrated interaction. In order to observe the changes in the state of the meter caused by the measurement interaction, it is necessary to initialize the meter in a well-defined state before each measurement. The maximal sensitivity is achieved when the meter is initialized in a pure state $\ket{\Phi}_M$. It is then possible to describe the changes in the state of the meter by unitary transformations conditioned by the eigenvalues $A_a$ of the target observable $\hat{A}$. These conditional changes of the meter characterize the information transfer from the system to the meter in terms of the quantum mechanics of the meter. 

For an arbitrary system state $\ket{\Psi}_S$, the effects of the unitary operation describing the measurement interaction can be represented by an expansion in terms of the eigenstates $\ket{a}$ of $\hat{A}$,
\begin{equation}
\label{eq:Mforce}
\hat{U}_{SM} \ket{\Psi}_S \otimes \ket{\Phi}_M = \sum_a \braket{a|\Psi}_S \ket{a} \otimes \exp \left( - \frac{i}{\hbar} g A_a \hat{B} \right) \ket{\Phi}_M.
\end{equation}
It should be noted that the expansion of the target observable $\hat{A}$ into eigenstates can be applied to any spectrum of eigenvalues $\{ A_a \}$. In the case of continuous variables, the sum should be replaced with an integral. In all cases, each eigenvalue corresponds to a specific unitary transformation acting on the meter state. We can therefore conclude that the information about $\hat{A}$ transferred to the meter in the measurement interaction is encoded in the form of a parameter $\varphi_B=g A_a$ that expresses the magnitude of the changes in the meter state. Although the actual values of $\varphi_B$ appear to be constraint by the specific set of eigenvalues $\{ A_a \}$ it is not necessary to know the eigenvalues in order to evaluate the change of the meter. In fact, the possibility of quantum interferences between different eigenvalues $A_{a1}$ and $A_{a2}$ indicates that a premature assignment of specific values to the magnitude of change expressed by the parameter $\varphi_B$ might distort the evaluation of the measurement result. In the spirit of \cite{vKam88} we therefore propose to consider the general dependence of the meter state $\ket{\phi(\varphi_B)}_M$ after the interaction on the unknown force $\varphi_B$ exerted by the system during the interaction without any assumptions about the available eigenvalues. The effect of the unknown force $\varphi_B$ is then represented by the unitary transformation of the initial meter state given by 
\begin{equation}
\label{eq:response}
\ket{\phi(\varphi_B)}_M = \exp \left( - \frac{i}{\hbar} \varphi_B \hat{B} \right) \ket{\Phi}_M.
\end{equation}
In the final stage of the measurement, the information encoded in this parameterized state needs to be extracted using an appropriate readout method. Since the value of $\varphi_B$ is not encoded in the form of an eigenvalue of a meter observable, the readout takes the form of a parameter estimation problem, where the target observable appears as a classical parameter in the unitary transformation of the quantum meter. Here, we will not make use of any prior information regarding the possible eigenvalues of $\hat{A}$. Instead, the magnitude of $\varphi_B$ will be deduced entirely from the observable changes in the meter state described by the parameter dependence of $\ket{\phi(\varphi_B)}_M$. 

\begin{figure}[t]
\begin{picture}(500,250)
\put(0,0){\makebox(500,250){
\scalebox{0.45}[0.45]{
\includegraphics{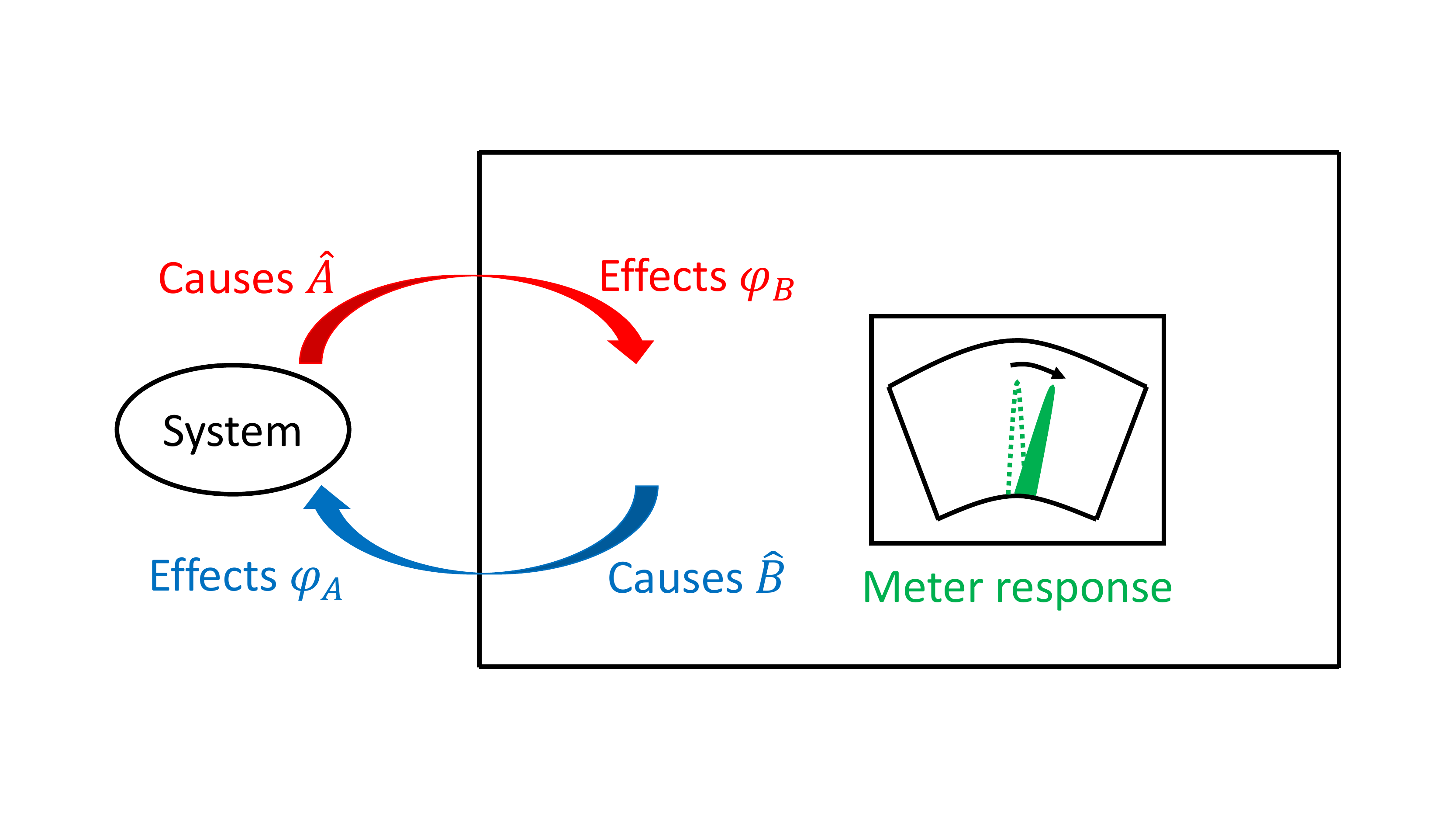}}}}
\end{picture}
\caption{\label{fig1}
Separation of the measurement interaction into causes and effects. The effect of the system observable $\hat{A}$ appears as an observable change of the meter system, where the magnitude of the change can be expressed by a parameter $\varphi_{B}$. Irrespective of the size or structure of the meter, the ability to distinguish different values of $\varphi_{B}$ depends on quantum uncertainties in the back action on the system described by the parameter $\varphi_A=g B_b$. 
}
\end{figure}

Although Eq.(\ref{eq:Mforce}) is a complete description of the measurement interaction, its formulation  emphasizes the effects of the system on the meter, potentially obscuring the corresponding effect of the meter on the system. This back action effect describes the unavoidable disturbance of the system state by the measurement interaction. Since the present analysis focuses on the characteristics of the meter, it is convenient to describe this back action effect by expanding the state of the meter in terms of the eigenstates $\ket{b}$ of the observable $\hat{B}$. The back action can then be represented by conditional unitary operations acting on the system,
\begin{equation}
\label{eq:Sforce}
\hat{U}_{SM} \ket{\Psi}_S \otimes \ket{\Phi}_M = \sum_b \exp \left( - \frac{i}{\hbar} g B_b \hat{A} \right) \ket{\Psi}_S \otimes \braket{b|\Phi}_M \ket{b}.
\end{equation}
This representation of the unitary transformation $\hat{U}_{SM}$ provides a more accessible description of the changes in the system caused by the interaction. In particular, it traces the disturbance of the system back to its origin in the uncertainties of the meter observable $\hat{B}$. Fig. \ref{fig1} illustrates the two aspects of the measurement interaction represented by the conditional unitary transformations in Eqs. (\ref{eq:Mforce}) and (\ref{eq:Sforce}). From the perspective of the observer outside the system, the target observable $\hat{A}$ is observed as a shift of the meter represented by the parameter $\varphi_B$ of the conditional unitary acting on the meter state. From the perspective of the system, the meter observable $\hat{B}$ that serves as the generator of the meter dynamics determines the magnitude of the force acting on the system, where the target observable $\hat{A}$ serves as the generator of the changes to the original system state. Different from the conditional force on the meter, this force serves no useful purpose in the measurement process. If the value of $\hat{B}$ were known, it would be possible to undo the changes to the system state $\ket{\Psi}_S$ by a corresponding local unitary acting on the system after the interaction \cite{Hof21}. However, the value of $\hat{B}$ is quantum mechanically uncertain, as given by the initial meter state $\ket{\Phi}_M$. Effectively, the system experiences a randomized unitary transformation, resulting in a mixed state output $\hat{\rho}_S(\mbox{out})$ with reduced coherences between the eigenstates of the target observable $\hat{A}$. The effects of the back action on the state of the system can therefore be characterized by the decoherence observed in off-diagonal elements of the density matrix in the eigenstate basis of $\hat{A}$,  
\begin{equation}
\label{eq:dephase}
\bra{a_1} \hat{\rho}_S(\mbox{out}) \ket{a_2}_S = \left( \sum_b |\braket{b|\Phi}_M|^2 \exp \left( - \frac{i}{\hbar} g  B_b \left( A_{a_1} - A_{a_2} \right)\right)\right) \braket{a_1 | \Psi} \braket{\Psi | a_2}_S.
\end{equation}
This description of decoherence corresponds to a standard open systems representation of the decoherence caused by external fluctuations. Since the decoherence is caused by unitary transformations generated by the observable $\hat{A}$ the amount of decoherence between two eigenstates depends on the difference between the eigenvalues of the two eigenstates, $A_{a_1}$ and $A_{a_2}$. Otherwise, it is entirely determined by the statistics of the meter eigenvalues $B_b$ given by the probabilities $|\braket{b|\Phi}_M|^2$ in the initial meter state $\ket{\Phi}_M$. It is therefore possible to characterize the disturbance of the system by the measurement interaction in terms of the statistics of the meter observable $\hat{B}$, independent of the initial state of the system $\ket{\Psi}_S$ or the possible values of the target observable $\hat{A}$.

The analysis presented above provides us with separate descriptions of the measurement signal received by the meter and the back action randomizing the state of the system. The relation between the two descriptions is provided by the meter observable $\hat{B}$, which serves as the generator of the changes to the meter state that represent the measurement signal. In the description of the back action, the same observable appears as a random parameter of the changes to the system state, where the amount of randomness in the value of $\hat{B}$ is the cause of the decoherence in the system. In the following, we will show how the laws of quantum mechanics establish a fundamental relation between these two roles of the meter observable $\hat{B}$.

\section{Analysis of meter sensitivity} 
\label{sec:sensitivity}

The meter response to the force exerted by the system is initially described by the parameterized quantum state given in Eq.(\ref{eq:response}). At this point the meter response is still fully quantum mechanical and no classical record of the measurement outcome exists. Independent of the precise procedure used to produce a measurement outcome, the causality relation between the output state of the meter and the measurement outcome $m$ must be represented by a positive operator valued measure (POVM) $\{\hat{E}(m)\}$ acting on the Hilbert space of the meter. Ideally, this POVM corresponds to a projection onto a measurement basis $\{ \mid m \rangle \}$ in a sufficiently large extended Hilbert space. The measurement information about the magnitude of the parameter $\varphi_B$ is then expressed by the conditional probabilities of the outcomes $m$ given by
\begin{equation}
\label{eq:probability}
P(m|\varphi_B) = |\braket{m|\phi(\varphi_B)}|^2.
\end{equation}
Different values of $\varphi_B$ are distinguished by the different probabilities of the outcomes $m$ conditioned by the interaction between the system and the meter. The ability to distinguish between two different values of the parameter depends on the statistical distance between the two conditional probabilities. As was shown in \cite{Pat19}, a particularly convenient statistical distance for the evaluation measurement resolution is the squared Hellinger distance. In the present context, we can define the resolution between a parameter value of $\varphi_B$ and a different parameter value of $\varphi_B + \varepsilon$ using the squared Hellinger distance
\begin{equation}
\label{eq:hellinger}
R(\varphi_B,\varphi_B+\varepsilon) = \frac{1}{2} \sum_m \left( \sqrt{P(m|\varphi_B+\varepsilon)} - \sqrt{P(m|\varphi_B)} \right)^2.
\end{equation}
A resolution much smaller than one indicates that the outcome probabilities $P(m|\varphi_B)$ and $P(m|\varphi_B + \varepsilon)$ are nearly indistinguishable. Even though the best estimate of the value of $\varphi_B$ will still depend on the precise outcome of $m$, the estimate will only be reliable if the parameter difference $\varepsilon$ is sufficiently large to produce a non-negligible resolution. By introducing appropriate criteria for the evaluation of the dependence of the resolution $R(\varphi_B,\varphi_B+\varepsilon)$ on the parameter difference $\varepsilon$, it is possible to define a quantitative resolution of $\delta \varepsilon$ describing the magnitude of the parameter difference $\varepsilon$ that can be resolved based on a single measurement outcome $m$.

We would like to emphasize that the present approach does not impose any limitations on the Hilbert space of the meter, on the initial state $\ket{\Phi}_M$ or on the operator $\hat{B}$. In general, the evaluation of the resolution $R(\varphi_B,\varphi_B+\varepsilon)$ from the effects of the unitary operation generated by $\hat{B}$ on the initial state $\ket{\Phi}_M$ can be quite complicated and the selection of an appropriate read out is not always obvious. It may still be useful to consider a particularly simple example to illustrate how the evaluation works. Perhaps the most simple example is that of a single qubit used as a meter, with quantum gate operations realizing the controllable interactions between the qubit meter and the system. This kind of meter system has already been realized in a number of experiments inspired by quantum information technologies. Notable examples include two level atoms used to measure fields in a cavity \cite{Rai01}, photon polarization \cite{Pry05}, photon paths in an interferometer \cite{Iin11}, and an artificial two level systems defined by neutron energies \cite{Gep18}.  In all of these cases, the meter generator $\hat{B}$ can be represented by an effective spin operator with eigenvalues of $\pm \hbar/2$, so that the unitary operation in Eq.(\ref{eq:response}) describes rotations of the Bloch vector by angles of $\varphi_B$ around the axis defined by $\hat{B}$. If the initial state is an equal superposition of the eigenstates of $\hat{B}$, a meter read out of spin components at an angle of $\alpha$ with respect to the orientation of the initial state results in probabilities of 
\begin{eqnarray}
\label{eq:qubit}
P(m=1) &=& \left(\cos(\frac{\alpha+\varphi_B}{2})\right)^2 \nonumber \\
P(m=2) &=& \left(\sin(\frac{\alpha+\varphi_B}{2})\right)^2. 
\end{eqnarray} 
In this case, the resolution given by the squared Hellinger distance is comparatively easy to calculate because of the trigonometric relations between the angles. The complete characterization of the qubit meter resolution is given by
\begin{equation}
\label{eq:Rsample}
R(\varphi_B,\varphi_B+\varepsilon) = 1 - \left|\cos\left(\frac{\varepsilon}{2}\right)\right|.
\end{equation}
The dependence of resolution on the parameter difference $\varepsilon$ for a qubit meter is periodic in $\varepsilon$ because of the limitations imposed by the two dimensional Hilbert space. This sets qubit meters apart from meters with continuous Hilbert spaces where the linear pointer shift caused by $\varphi_B$ usually results in a gradual increase of the resolution towards a nearly constant resolution of one at large values of $\varepsilon$. It is therefore important to identify universal characteristics of the resolution that can be used to compare the performances of different meters in a meaningful way. 

In order to classify the large variety of possible resolution functions $R((\varphi_B,\varphi_B+\varepsilon)$ in terms of a single characteristic quantitative resolution $\delta \varepsilon$ it is useful to consider the Taylor expansion of the resolution for small parameter differences $\varepsilon$. Since the resolution is zero at $\varepsilon=0$ and the resolution is positive by definition, the approximate resolution at sufficiently small values of $\varepsilon$ is given by the quadratic term of this Taylor expansion,
\begin{equation}
\label{eq:approx}
R(\varphi_B,\varphi_B+\varepsilon) \approx \frac{1}{2} \left( \left.\frac{\mathrm{d}^2}{\mathrm{d} \varepsilon^2} R(\varphi_B , \varphi_B+\varepsilon) \right |_{\varepsilon = 0} \right) \varepsilon^2.
\end{equation}
Since the quadratic term dominates the $\varepsilon$ dependence of the resolution at low parameter differences $\varepsilon$, it provides a reliable estimate of the parameter differences necessary to achieve non-negligible resolution values. Specifically, our goal is to estimate the region of parameters that cannot be distinguished in a single measurement. The quantitative resolution $\delta \varepsilon$ should therefore be defined in such a way that the resolution can be expected to change from values much smaller than one to values close to one around the parameter difference of $\delta \varepsilon$. Based on the Taylor expansion in Eq.(\ref{eq:approx}) we can estimate that the resolution will increase to a value of $R \approx 1/8$ as the parameter difference increases from zero to a resolution value given by 
\begin{equation}
\label{eq:quantify}
\delta \varepsilon =  \frac{1}{2 \sqrt{\left. \frac{\mathrm{d}^2}{\mathrm{d}\varepsilon^2} R(\varphi_B, \varphi_B + \varepsilon) \right|_{\varepsilon = 0}}}.
\end{equation}
We can thus identify a range of parameter differences with $|\varepsilon|< \delta \varepsilon$ for which the estimated resolution between $\varphi_B$ and $\varphi_B + \varepsilon$ is expected to be below one eighth. The Taylor expansion also shows that the resolution is expected to rise quickly to values higher than one eighth as the parameter difference increases. Although the precise form of the transition to resolutions close to one will be modified by the higher order terms of the Taylor expansion, the quantitative resolution $\delta \varepsilon$ will thus be a good indicator of the minimal parameter difference needed to achieve resolutions close to one. 

The example of a qubit meter given in Eqs. (\ref{eq:qubit}) and (\ref{eq:Rsample}) can be used to illustrate the differential characterization of the resolution. The second derivative of the resolution in Eq.(\ref{eq:Rsample}) at $\varphi_B=0$ gives a value of $1/4$, resulting in a quantitative resolution measure of $\delta \varepsilon = 1$. The qubit meter defined by the read out probabilities in Eq. (\ref{eq:qubit}) therefore represents a unit resolution in $\varphi_B$. Comparison of the Taylor expansion with the precise resolution at $\varepsilon=1$ shows that the precise resolution is about two percent lower than the value of 1/8 given by the Taylor expansion in Eq.(\ref{eq:approx}). The definition of quantitative resolution in Eq.(\ref{eq:quantify}) is therefore sufficiently accurate for comparisons between qubit meters and other types of meters.  

Eq.(\ref{eq:quantify}) relates the quantitative resolution of a single measurement outcome to the second derivative of the resolution at a parameter difference of $\varepsilon=0$. This second derivative can be expressed in terms of the differential dependence of the conditional probabilities $P(m|\varphi_B)$ on the parameter $\varphi_B$,
\begin{equation}
\label{eq:sensitivity}
\left. \frac{\mathrm{d}^2}{\mathrm{d}\varepsilon^2} R(\varphi_B , \varphi_B + \varepsilon) \right|_{\varepsilon = 0} = \frac{1}{4} \sum_m \frac{1}{P(m|\varphi_B)} \left( \frac{\mathrm{d}}{\mathrm{d}\varphi_B} P(m|\varphi_B) \right)^2.
\end{equation}
Interestingly, the sum over measurement outcomes $m$ on the right hand side of the equation is equal to the Fisher information of the parameterized probability distribution $P(m|\varphi_B)$ \cite{Fis25},
\begin{equation}
\label{eq:Finformation}
F = \sum_m \frac{1}{P(m|\varphi_B)} \left( \frac{\mathrm{d}}{\mathrm{d}\varphi_B} P(m|\varphi_B) \right)^2.
\end{equation}
It may be worth noting that this result relates the Fisher information to the resolution of a finite difference between the values of the parameter $\varphi_B$ using only a single measurement outcome $m$. Although parameter estimation theory could be applied to the meter readout scenario, the present analysis of statistical distances makes no reference to the optimization of an estimate of $\varphi_B$ based on the outcome $m$. Instead the Fisher information provides an estimate of the range of parameter differences $\varepsilon$ at which it is effectively impossible to distinguish the statistics of outcomes $m$ associated with the parameter $\varphi_B$ from the statistics of outcomes $m$ associated with a parameter of $\varphi_B+\varepsilon$. It thus serves as an estimate of a non-vanishing statistical distance between the different values of the parameter $\varphi_B$.

The fact that the quantitative resolution $\delta \varepsilon$ is given by the Fisher information $F$ of the parameterized probability distribution $P(m|\varphi_B)$ simplifies the identification of the quantum mechanical limits of meter sensitivities. Since the parameter $\varphi_B$ is the parameter of a unitary transformation generated by the physical property $\hat{B}$ of the meter system, the upper bound of the Fisher information associated with any possible measurement $\{\hat{E}(m)\}$ of the meter is given by the uncertainty of the generator $\hat{B}$. Specifically, the well known upper bound of the Fisher information corresponds to the Mandelstam-Tamm uncertainty relation between the parameter of a unitary and the generator \cite{Bra94,Luo00,Luo04},
\begin{equation}
\label{eq:QCRB}
F \le \frac{4}{\hbar^2} \Delta B^2.
\end{equation}
We can now apply this well known upper bound of the Fisher information to determine the upper bound of the meter sensitivity. Here, we define the sensitivity of the meter as the inverse of the quantitative resolution $\delta \varepsilon$. Its upper bound is given by
\begin{equation}
\label{eq:limit}
\frac{1}{\delta \varepsilon} \le \frac{2}{\hbar} \Delta B.
\end{equation}
The quantitative resolution $\delta \varepsilon$ represents the amount of force needed to change the meter state in a noticeable manner. Eq. (\ref{eq:limit}) shows that the corresponding parameter difference of $\varepsilon$ can only be resolved by a meter with a minimal generator uncertainty of $\hbar/(2 \varepsilon)$. This requirement of a quantum uncertainty in the meter is completely independent of the type of meter or its precise initialization. In particular, it is not possible to construct a ``classical'' meter that achieves its sensitivity to the system without any quantum uncertainties. Consistency between quantum mechanics and classical physics requires that the sensitivity of any meter system is determined by the amount of quantum coherence expressed by the Fisher information and the associated quantum uncertainty of the generator observable in the meter \cite{Luo17}. 

In the case of a qubit meter, the uncertainty of the generator is given by the definition of $\hat{B}$ as a spin-1/2 operator. For the equal superposition of eigenstates used as an initial meter state, this operator has an uncertainty of $\Delta B =\hbar/2$. The uncertainty in Eq.(\ref{eq:limit}) is therefore saturated by the quantitative resolution of $\delta \varepsilon = 1$ determined from the meter dynamics. This is the typical result for any optimized quantum meter. However, most realistic measurement devices will have much higher levels of uncertainty due to imperfections in the meter setup. The limit given here can only be achieved at maximal levels of control. However, the lower bound of the quantum coherence in the meter imposed by its sensitivity to the effects of the target observable $\varphi_B$ is a universal law of physics that requires the presence of a minimal amount of quantum coherent effects in any meter, no matter how macroscopic or noisy the actual setup might be. 

\section{Effects of meter uncertainties on the quantum state of the system}
\label{sec:system}

As shown by Eq.(\ref{eq:Sforce}) the magnitude of the change in the state of the system during the measurement interaction is given by the eigenvalues $B_b$ of the meter generator $\hat{B}$. The statistical distribution of the values of $B_b$ given by the meter probabilities $|\braket{b|\Phi}_M|^2$ results in a reduction of coherences between different eigenstates of the target observable $\hat{A}$ in the system as described by Eq.(\ref{eq:dephase}). The amount of decoherence can be defined as the reduction of the off-diagonal elements,
\begin{equation}
\label{eq:Def}
D(a_1,a_2)= 1- \left|\frac{\bra{a_1} \hat{\rho}_S(\mbox{out}) \ket{a_2}}{\braket{a_1 | \Psi} \braket{\Psi | a_2}}\right|.
\end{equation}
The precise dependence of the decoherence on the eigenvalues $A_1$ and $A_2$ and the statistics of the meter generator $\hat{B}$ is given by Eq.(\ref{eq:dephase}). The dependence of the decoherence $D(a_1,a_2)$ on the difference between the eigenvalues $A_1$ and $A_2$ can be summarized using the difference $\varepsilon$ between the corresponding parameters $\varphi_B$ and $\varphi_B+\varepsilon$, resulting in a function that depends only on the parameter difference $\varepsilon$, 
\begin{equation} 
\label{eq:parameterD}
D(\varepsilon) = 1-\left| \sum_b |\braket{b|\Phi}_M|^2 \exp \left( - \frac{i}{\hbar} B_b \varepsilon \right) \right|.
\end{equation}
The dependence of the decoherence on the parameter difference $\varepsilon$ is completely defined by the statistical distribution of generator eigenvalues $B_b$ in the state of the meter.  In general, the decoherence is zero at $\varepsilon=0$ and increases gradually as the difference $\varepsilon$ between the parameters of the parameters increases. For the qubit meter example, 
\begin{equation} 
\label{eq:Dsample}
D(\varepsilon) = 1-\left|\cos\left(\frac{\varepsilon}{2}\right)\right|.
\end{equation}
As shown \cite{Pat19}, the decoherence must always be greater than or equal to the resolution. Comparison of Eq.(\ref{eq:Rsample}) and Eq.(\ref{eq:Dsample}) show that the qubit meters saturate this bound. 

In close analogy to the definition of the quantitative resolution $\delta \varepsilon$ we can use the Taylor expansion of the decoherence in $\varepsilon$ to estimate the values of $\varepsilon$ for which the decoherence between the eigenstates of $\hat{A}$ is still negligible. Since decoherence applies to the system and its properties, it is convenient to quantify the result in terms of the decoherence free range of eigenvalue differences $A_{a_1} - A_{a_2}$. This range of decoherence free eigenvalue differences can be estimated using the Taylor expansion of the decoherence $D(\varepsilon)$. If the decoherence free distance $C_A$ is defined as
\begin{equation}
\label{eq:fdf}
C_A =  \frac{1}{2 g \sqrt{\left. \frac{\mathrm{d}^2}{\mathrm{d}\varepsilon^2} D(\varepsilon) \right|_{\varepsilon = 0}}},
\end{equation}
we can expect a decoherence below $1/8$ for eigenvalue differences with $|A_{a_1}-A_{a_2}|<C_A$. The decoherence free distance is therefore a good measure of the maximal coherence length in the target observable $\hat{A}$ after the measurement interaction. 

Since $D(\varepsilon)$ is completely determined by the statistics of $\hat{B}$ in the meter state it is not surprising to find that the decoherence free distance $C_A$ is completely determined by the uncertainty $\Delta B$ of the meter generator $\hat{B}$. The coherence length maintained by the measurement interaction is proportional to the inverse of the generator uncertainty,
\begin{equation}
\label{eq:coherence}
C_A = \frac{\hbar}{2 g \Delta B}. 
\end{equation}
This deterministic relation between system decoherence and meter uncertainty ensures that the meter sensitivity limits the coherence of the system after the measurement, eliminating any coherences between eigenstates that can be distinguished from each other in the measurement. For a proper evaluation in terms of the target observable $\hat{A}$ the quantitative resolution needs to be converted into the resolution of $\hat{A}$ given by $\delta A=\delta \varepsilon/g$. According to Eq.(\ref{eq:limit}) the generator uncertainty required for a resolution of $\delta A$ is given by
\begin{equation}
\label{eq:Rlimit}
\delta A \ge \frac{\hbar}{2 g \Delta B}. 
\end{equation}
This inequality ensures that the resolution $\delta A$ is limited to eigenvalue differences larger than the decoherence free distance $C_A$ given in Eq. (\ref{eq:coherence}). Our analysis thus shows in a quantitative manner how the distinguishability of effects of the target observable $\hat{A}$ on the meter depends on the uncertainty of the back action of the meter on the system. There can be no observable effect of $\hat{A}$ on the outside world without a corresponding reduction of the decoherence free distance $C_A$. Significantly, the relation between the two effects is determined by the quantum mechanics of the external meter. It is therefore possible to trace the origin of quantum uncertainties to physical noise sources outside of the system. 

\section{External effects of a quantum system}
\label{sec:context}

The present analysis is based on the formal separation of the system and the meter using the complementary representations of the interaction given by $\hat{U}_{SM}$. The most important insight gained by this analysis is that the external effects of the target observable $\hat{A}$ are limited to the observable changes of the meter state caused by the magnitude of its specific value. As explained in the introduction, this limitation of the effects of an observable $\hat{A}$ is closely related to the incompatibility between different measurement contexts \cite{KS,Fra18,Bru18,Pro19}. An assignment of eigenvalues to the physical properties of a quantum system is only justified if there is an external effect of that physical property on the environment. The relation between the quantitative resolution $\delta A$ of the measurement and the coherence free distance $C_A$ therefore illustrates the gradual emergence of a specific measurement context as the interaction strength increases. This is similar to the complementarity between visibilities and which way information in a ``welcher Weg'' experiment \cite{Bjo98,Bar09}. As the eigenvalues of $\hat{A}$ are resolved, the coherence between the eigenstates must disappear. At the same time, the value of $\hat{A}$ is always encoded as the parameter of a conditional unitary transformation of the meter. It is important to ask in what sense this information about the value of $\hat{A}$ determines the observable reality of the physical property $\hat{A}$. If the measurement is fully resolved, the projective measurement limit is achieved and Born's rule applies to the measurement outcomes. However, any additional information about other physical properties is eliminated by the random dynamics of the measurement back action. The formalism seems to encode this loss of information in the decoherence of the system state. However, the analysis above has shown that this loss of information is actually determined by the physical action of a force on the system necessarily exerted by the meter. 

In the weak measurement limit, information about other physical properties is maintained and a different measurement context can still be selected by a final measurement of the unchanged system state. This is the reason why weak values can be understood as a valid representation of the context dependence of physical properties \cite{Tol07,Dre10,Hal16,Hof20,Bud21}. The criticism that weak values actually represent quantum interference effects is somewhat misleading since quantum coherence is merely a mathematical representation of the relations between different measurement contexts \cite{Dre15,Sok16}. The real problem is that quantum state components cannot possibly describe measurement independent realities inside the system. The assumption that Born's rule can be applied inside a system cannot be applied consistently as shown in \cite{Fra18,Bru18}. Since Born's rule emerges naturally in the limit of fully resolved measurements, it seems to be more reasonable to limit the discussion of reality to the information available outside of the system. The present analysis has shown that this information can only be obtained by physically randomizing the dynamics of the system. This randomization is quantum coherent because the external meter depends on quantum coherences in its microscopic degrees of freedom for its macroscopic sensitivity. Measurement contextuality is mediated by the quantum nature of interactions between a system and its environment. It is physically impossible to directly access any internal reality of the system, but the determinism of quantum interactions always ensures that the actual information about a system available in its environment is consistent with itself.

\section{Conclusions}
\label{sec:conclude}

In conclusion, we have shown that the effects of a system on the meter used in a quantum measurement are limited by the quantum coherence available in the meter. This quantum coherence is the origin of the back action that erases the information about other physical properties of the system encoded by quantum coherences between the eigenstates of the target observable $\hat{A}$. The quantitative resolution $\delta A$ is achieved by a back action uncertainty that limits the decoherence free distance $C_A$ to a maximal value equal to the resolution. The consistency of quantum mechanics is therefore ensured by the presence of quantum coherence in all physical systems, including the macroscopic elements of an experimental setup. Although questions remain about the macroscopic selection of a specific measurement outcome, the analysis above has shown that the decoherence caused by a measurement back action can always be traced to quantum coherent noise sources in the environment of the system. There is always a deterministic explanation of the origin of quantum uncertainties. Superpositions and their quantum coherences merely encode the causality of quantum measurements described by unitary transformations.

\end{document}